\documentclass[twocolumn,showpacs,preprintnumbers,pra,nofootinbib,superscriptaddress]{revtex4}

\usepackage[english]{babel} 	
\usepackage{graphicx} 			
\usepackage{amssymb} 			
\usepackage{amsthm} 			
\usepackage{amsmath} 			
\usepackage{siunitx} 			
\usepackage{subfigure} 			
\usepackage{epstopdf}

\begin{document}


\title{Analytic solutions of topologically disjoint systems}

\author{J.~R. Armstrong}
\affiliation{Department of Physics, Winona State University, Winona, MN 55987, USA } 

\author{A.~G. Volosniev, D.~V. Fedorov, A.~S. Jensen, N.~T. Zinner}
\affiliation{Department of Physics and Astronomy, Aarhus University, DK-8000 Aarhus C, Denmark}

\date{\today}


\begin{abstract}
We describe a procedure to solve an up to $2N$ problem where the particles
are separated topologically in $N$ groups with at most two particles
in each.  Arbitrary interactions are allowed between the (two)
particles within one group.  All other interactions are approximated
by harmonic oscillator potentials. The problem is first reduced to an
analytically solvable $N$-body problem and $N$ independent two-body
problems.  We calculate analytically spectra, wave functions, and
normal modes for both the inverse square and delta-function two-body
interactions.  In particular, we calculate separation energies between two strings of particles.  
We find that the string separation energy increases with $N$ and interaction strength.  
\end{abstract}


\maketitle

\section{Introduction}
Analytic models were unavoidable in physics before the advent of
computers.  This forced researchers to extract the essence of the
problems and design the models to catch the crucial properties.
Appropriately done these simplifications delivered deep insight and
the necessary realistic results.  Computers are extremely useful to
provide solutions of more and more complicated problems.  Advanced
research is then almost by definition forced to push implementations to the
limit of the available computer capacity.

The increased complexity of problems combined with large numerical
calculations require larger efforts to understand and develope an
intuition for the underlying physics.  Getting the deeper insight is
more difficult for complicated problems, but nevertheless very
desirable for several reasons.  Beside the direct usefulness of basic
understanding, it is also efficient in construction of improved
algorithms, which in turn would allow investigation of even more
complicated problems.

We recapitulate briefly a number of important reasons to employ
suitable analytic methods.  First, model construction must concentrate
on the crucial features, and thus formulate and order the issues
according to importance.  Second, in many perturbative
approaches in physics the potentials are expanded to second
order in coordinates or momenta, and the emerging harmonic oscillator
problems are analytic.  Third, a problem can be beyond even present
day computer capacity.  Fourth, advanced numerics can be greatly
improved by combination with analytic tools and insight.  Fifth,
analytic models can hardly be overestimated as a tool in teaching on
all levels.  Sixth, present experimental frontline research on cold
atomic gases employ unprecedentedly simple (almost) analytically
solvable potentials \cite{bloc08}.

The harmonic oscillator is the simplest and most useful analytic
potential at our disposal and this 
has been exploited throughout the history of physics. A relevant 
example dating all the way back to 1926 is by Werner 
Heisenberg who used harmonic Hamiltonians to gain insight into 
the many-body problem \cite{heisenberg1926} (within a year of him 
inventing the new quantum theory). Later on the harmonic
oscillator was instrumental in providing insights for models of the
atomic nucleus \cite{goep55}. Harmonic interactions have also been 
used as a replacement for Coulomb potentials in atoms in order to 
gain analytical insights (the so-called Moshinsky atom or pseudoatom)
\cite{moshinsky1968,moshinsky1985}. 
These models continue to produce new
insights into aspects of atomic systems such as the density matrix \cite{amovilli2003,schilling2013a}, 
correlations \cite{kestner1962,taut1993,lopez2006a,lopez2006b,laguna2011} and entropy \cite{amovilli2004}, density functionals \cite{riveros2012a}, and 
entanglement properties \cite{pipek2009,yanez2010,bouvrie2012,riveros2012b,koscik2013,riveros2014,bouvrie2014}.
Harmonic models have been a subject of great interest in several other
fields including quantum dots \cite{johnson1991}, quantum statistics \cite{liang2012,schilling2013b}, 
black holes thermodynamics \cite{bombelli1986,srednicki1993}, and recently also area laws and 
entanglement in quantum many-body systems \cite{eisert2010}. 

In the field of cold atomic gases the harmonic oscillator 
method has also served as an important model to gain insights.
For instance, it has been used extensively as a starting 
point for path integral calculations of quantum gas properties
\cite{bros97,zalu00,yan03,gajd06,tempere1997}.
More recently, an $N$-body model for harmonically interacting 
particles in an external harmonic confining potential has been
solved \cite{arms11,arms12}, and the spectra have been used to 
investigate thermodynamics and virial expansions \cite{arms12a,arms12b}.
All these studies attest to the versatility of the harmonic approach 
as an analytical tool that can address even advanced frontline 
research questions.

A recent frontier in ultracold 
atomic gases is the creation of ultracold polar molecules \cite{lahaye2009,carr2009}
with long-range interactions that hold great promise for greating some 
unique exotic quantum systems \cite{carr2009,baranov2008}. To 
avoid strong dipolar loss due to the attractive head-to-tail interactions,
the dipoles should be confined to lower dimensional geometries such as 
tubes or layers, and a layered system of dipoles was recently experimentally
realized \cite{mira11}. In such a system one expects the formation of 
chains of dipoles which are bound structures across several layers \cite{wang06,arms-dip12,barbara2011,arms-dip13}. 
Similarly, in one-dimensional tubes one also expects this bound state
formation \cite{klawunn2010,barbara2011,artem-dip2013}. This chain formation 
takes place in a topologically disjoint system as the layers or tubes 
are physically disconnected when tunneling between them is negligible. 
However, they still provide a very interesting physical system as the 
dipolar interactions are long-range and act between physically 
disconnected parts of the system. Here we consider 
some general models that allow an analytical approach to such topologically
disjoint geometries. 

More generally, 
the purpose of the present paper is to discuss an
analytic model which reduces a $2N$ problem in a non-trivial 
geometry with disjoint regions to an approximate but
analytic $N$-body problem and $N$ individual two-body problems. The
particles must pairwise have the same masses but the in-pair
interaction can be arbitrary.  The same four two-body oscillator
interactions are preferred between two such pairs.  Otherwise there
are no constraints on masses and arbitrary oscillator intereractions
are allowed. A seperable model like the one we discuss here has been 
studied recently within the context of exactly solvable models
\cite{karwowski2004,karwowski2008} and applied as a 
model of two-electron atoms and molecules \cite{karwowski2010}. In the 
current work we will be applying this kind of Hamiltonian to other 
model systems that are of relevance for neutral atoms in cold atomic
gases.

The paper is organized as follows. In Section~\ref{method} we derive the
conditions for the huge simplification obtained by decoupling the
different degrees of freedom.  In section~\ref{analytics} we specialize to
analytic two-body problems, and in Section~\ref{numerics} we present a number of
corresponding examples for illustrations.  Finally in Section~\ref{conclusions} , we
briefly summarize, conclude and give perspectives.

\section{Method}\label{method}
We consider a system of particles which can be divided into $N$ groups
where each consists of one or two particles.  A priori, all particles
may be identical or different, and any spatial dimension is allowed.
The two-body interactions between all the topologically separated groups of particles are assumed
to be harmonic oscillators.  In addition, we allow completely
arbitrary interactions between the pairs of particles within each of
the $N$ groups.  Identical particles must obey Fermi or Bose symmetry
relations, but the group distinction must be maintained.  The perhaps
artificially appearing grouping and the related conditions arise from
anticipating a strongly simplifying decoupling of the many degrees of
freedom.  The corresponding additional decoupling conditions are
derived in the first subsection, and specific features of the
solutions are discussed in the second subsection.

\begin{figure}
\centering
{\includegraphics[width=0.95\columnwidth]{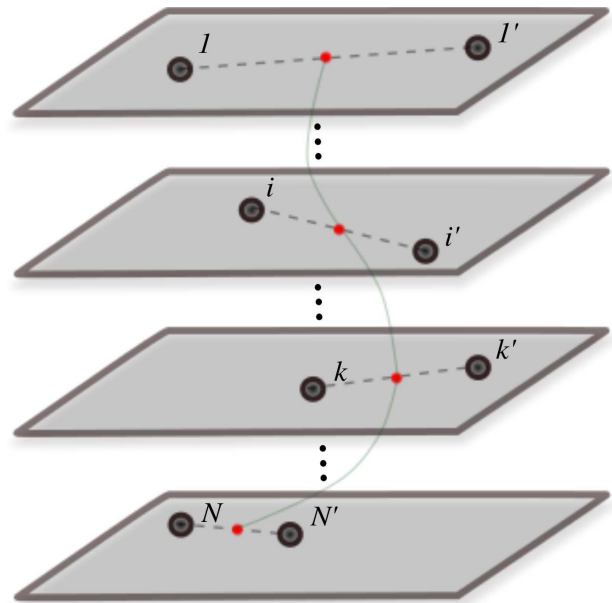}}
\caption{(Color online) Schematic picture of a possible $2D$ layer
  geometry with two particles in each layer (for instance $i$ and $i'$). The
  pairwise intra-layer relative motion are independent of other pairs,
  and as well independent of the motion of their intra-layer
  center-of-masses.  The particles could be constrained to stay on a
  one-dimensional curve within each layer reducing the problem to
  $1D$.  The particles may also be allowed to move freely in all three
  dimensions, but for the decoupling to work we insist that only $i-i'$ pairs can have
  arbitrary relative two-body interactions.
\label{fig0}}
\end{figure}

The required distinction between particles and interactions can be
achieved by geometrically separated systems.  The most obvious
possibilities are series of separate two-dimensional layers or
one-dimensional tubes where each layer or tube is occupied by one or
two particles.  We provide a visualization in fig.~\ref{fig0}.  The
inter-layer and/or inter-tube two-body interactions then must be of
harmonic oscillator form whereas arbitrary intra-layer and/or
intra-tube two-body interactions are allowed.  In addition one-body
harmonic oscillator potentials acting on each of the particles are
also allowed.  The restrictive fundamental conditions may even be
possible in some three-dimensional structures.  Imagine a lattice with
up to two particles per site with arbitrary on-site interactions, and
harmonic potentials between particles on different sites.

\subsection{Hamiltonian}
The $N$ groups are labeled by $k=1,2,3,...,N$ and the two particles in
each group are labeled by $k$ and $k'$, where we first assume two
particles in all $N$ groups.  To simplify the discussion we shall
refer to the system as a series of $N$ doubly occupied layers.  Only
$k$ and $k'$ are allowed arbitrary interactions, while two-body individual harmonic oscillator potentials are assumed for any
of the other interactions.  The Hamiltonian, $H$, of this $N$-layer
system is then
\begin{eqnarray}
&&H=-\frac{\hbar^2}{2}\sum_{k=1}^N\left(\frac{1}{m_k}\nabla_k^2
+\frac{1}{m_{k'}}\nabla_{k'}^2\right) + V^{(shift)} \nonumber \\ 
&&+\sum_{k=1}^NV_k(\mathbf{r}_k-\mathbf{r}_{k'})+
\frac{1}{2}\sum_{k=1}^N \omega_{0k}^2 \left(m_kr_k^2+m_{k'}r_{k'}^2\right)
 \nonumber \\
 &&+\frac{1}{2}\sum_{i=1}^{N-1} \sum_{k=i+1}^{N}
\left[\mu_{ik}\omega_{ik}^2\left(\mathbf{r_i}
-\mathbf{r_k}\right)^2+\mu_{ik'}\omega_{ik'}^2(\mathbf{r_i}
-\mathbf{r_{k'}})^2\right.  \nonumber \\
&&\left.+\mu_{i'k}\omega_{i'k}^2(\mathbf{r_{i'}}-\mathbf{r_k})^2
+\mu_{i'k'}\omega_{i'k'}^2(\mathbf{r_{i'}}-\mathbf{r_{k'}})^2\right] \;,\label{eq1}  
\end{eqnarray}
where $\mathbf{r}_k$ and $m_k$ are coordinate and mass of particle
$k$, $\mu_{ik}=m_im_k/(m_i+m_k)$ is the reduced mass between particles
$i$ and $k$, $V_k$ is the intra-layer interaction between particles
$k$ and $k'$ in the $k$th layer, $\omega_{0k}=\omega_{0k'}$ is the
one-body external field frequency, $\omega_{ik}$ is the harmonic
frequency between the particles $i$ and $k$ in the layers $i$ and $k$,
and $V^{(shift)}$ is a constant zero-point energy shift which is needed to compare with experimental results.  
Numerically, $V^{(shift)}$ is the
sum of  the individual interactions, $v^{(shift)}_{ik}$, between
particles $i$ and $k$, that is
\begin{eqnarray} \label{shift0}
&&V^{(shift)} =  \nonumber\\ &&
\sum_{i=1}^{N-1} \sum_{k=i+1}^{N} \left(v^{(shift)}_{ik}+  
v^{(shift)}_{i'k} + v^{(shift)}_{ik'} + v^{(shift)}_{i'k'} \right)  .
\end{eqnarray}
which are chosen to fit the energies of the individual two-body systems.  All these definitions are completed by repeating with primes on each
of the indices as applied in the equations.  The different terms are
divided and explicitely written according to layers and thereby
reducing the summation index to run over layers.

We transform the coordinates in the different layers to relative,
$\mathbf{\tilde{r}_k}$, and center-of-mass two-body coordinates,
$\mathbf{R_k}$.  Defining $M_k=m_k+m_{k'}$, the transformations both
ways become
\begin{eqnarray} \label{tran1}
\mathbf{\tilde{r}_k}&=&\mathbf{r_k}-\mathbf{r_{k'}} \;;\;\;\;
\mathbf{R_k}=\frac{m_k\mathbf{r_k}+m_{k'}\mathbf{r_{k'}}}{M_k} \;, \\
\mathbf{r_k}&=&\mathbf{R_k}+\frac{m_{k'}}{M_k}\mathbf{\tilde{r}_k}  \;;\;\;\;
\mathbf{r_{k'}}=\mathbf{R_k}-\frac{m_k}{M_k}\mathbf{\tilde{r}_k} \;.\label{tran2}
\end{eqnarray}
Inserting these transformations into the Hamiltonian in
eq.~(\ref{eq1}) leads to many terms.  We consider first the one-body
terms from the kinetic energy operators and external fields, that is
\begin{eqnarray}  \nonumber
&&\frac{1}{2}\left(\omega_{0k}^2(m_kr_k^2+m_{k'}r_{k'}^2)
-\frac{\hbar^2}{m_k}\nabla_k^2 - \frac{\hbar^2}{m_{k'}}\nabla_{k'}^2 \right) = \\
&& \frac{1}{2}\left( \omega_{0k}^2 (M_kR_k^2+\mu_{kk'}\tilde{r}_k^2)
 -\frac{\hbar^2}{M_k}\nabla_{R_k}^2 - 
\frac{\hbar^2}{\mu_{kk}}\nabla_{\tilde{r}_k}^2\right)\;.
\end{eqnarray}

Then we transform the two-body oscillator terms in eq.~(\ref{eq1}) while assuming $\omega_{kk}=0$, $\omega_{ik}=\omega_{ki}$ and also for the primed indices.  We find
with the substitution in eq.~(\ref{tran2}) that
\begin{eqnarray}
&& \frac{1}{4}\sum_{i=1}^{N}  \sum_{k=1}^{N}  
[\mu_{ik}\omega_{ik}^2(\mathbf{r_i}-\mathbf{r_k})^2 
+\mu_{i'k}\omega_{i'k}^2(\mathbf{r_{i'}}-\mathbf{r_k})^2  \nonumber \\
 && + \mu_{ik'}\omega_{ik'}^2(\mathbf{r_i}-\mathbf{r_{k'}})^2 
 + \mu_{i'k'}\omega_{i'k'}^2(\mathbf{r_{i'}}-\mathbf{r_{k'}})^2]
 \nonumber \\
&& =  \frac{1}{4} \sum_{i=1}^{N}  \sum_{k=1}^{N}  (\mathbf{R_i}-\mathbf{R_k})^2  
 [\mu_{ik}\omega_{ik}^2  \nonumber \\
 &&  + \mu_{i'k}\omega_{i'k}^2 + 
 \mu_{ik'}\omega_{ik'}^2 + \mu_{i'k'}\omega_{i'k'}^2]
 \nonumber\\
&& + \frac{1}{2} \sum_{i=1}^{N}\mu_{ii'}  \tilde{r}_i^2  \sum_{k=1}^{N} 
 \left[ \frac{m_i}{M_i} \left(\frac{m_k}{m_{i'}+m_k}\omega_{i'k}^2 +  
\frac{m_{k'}}{m_{i'}+m_{k'}} \omega_{i'k'}^2 \right)\right. \nonumber \\ 
&& \left.+ \frac{m_{i'}}{M_i} \left(\frac{m_{k'}}{m_{i}+m_{k}} \omega_{ik}^2 +  
\frac{m_{k}}{m_{i}+m_{k'}} \omega_{ik'}^2 \right) \right]
 \nonumber \\ \nonumber
&& + \frac{1}{2} \sum_{i=1}^{N}  \sum_{k=1}^{N} 
 \mu_{ii'} \mu_{kk'}  \mathbf{\tilde{r}_i}\cdot\mathbf{\tilde{r}_k}
 \left[\frac{\omega_{ik'}^2  }{m_{i}+m_{k'}} - \frac{\omega_{ik}^2}{m_{i}+m_{k}}\right. 
 \\ \nonumber &&\left. + \frac{ \omega_{i'k}^2 }{m_{i'}+m_{k}}
-\frac{\omega_{i'k'}^2}{m_{i'}+m_{k'}}\right] \\ \nonumber
&& +  \sum_{i=1}^{N}  \sum_{k=1}^{N} 
\mu_{kk'} \mathbf{\tilde{r}_k}\cdot (\mathbf{R_k}-\mathbf{R_i}) 
   \left[ \frac{m_{i'}}{m_{i'}+m_{k}} \omega_{i'k}^2 \right.\\  \label{eq5} &&
 \left. - \frac{m_{i'}}{m_{i'}+m_{k'}} \omega_{i'k'}^2  +
 \frac{m_{i}}{m_{i}+m_{k}} \omega_{ik}^2 
 - \frac{m_{i}}{m_{i}+m_{k'}} \omega_{ik'}^2 \right] \;,
\end{eqnarray}
where the double summations independently run over all values of $i$
and $k$ from $1$ to $N$. Thus, to give the corresponding expressions
in eq.~(\ref{eq1}) we have multiplied by $1/4$.  All terms in the
Hamiltonian are now rewritten in terms of the
$(\mathbf{\tilde{r}_k},\mathbf{R_k})$ coordinates, since
$V_k(\mathbf{r_k}-\mathbf{r_k'}) = V_k(\mathbf{\tilde{r}_k})$.

We have so far assumed that each layer is occupied by precisely two
particles. It is a simplification if some layers only contain one
particle.  The equations are essentially still valid, provided we
insert $m_{k'}=0$ in the layers with only one particle.  The problem
arising in the kinetic energy operator is solved by disregarding the
corresponding infinite term.  The remaining formalism is then the
same.  The opposite direction of adding more particles in a layer is
much more complicated, since now nine terms appear by rewriting the
two-body interactions between two layers in Jacobi coordinates.

\subsection{Decoupling conditions}
The $2N$-body problem is still as complicated as in the formulation with
the original particle coordinates.  However, the coupling terms
between relative, $\mathbf{\tilde{r}_k}$, and center-of-mass coordinates, $\mathbf{R}_k$,vanish if
\begin{eqnarray} \label{cond1}
  && m_{i'}\left(\frac{\omega_{i'k'}^2}{m_{i'}+m_{k'}}
 - \frac{\omega_{i'k}^2}{m_{i'}+m_{k}}  \right)   \nonumber \\ && =   m_{i} \left(
\frac{\omega_{ik}^2}{m_{i}+m_{k}}-\frac{\omega_{ik'}^2}{m_{i}+m_{k'}}  \right) \;.
\end{eqnarray}
This reduces from $2N$ to two independent $N$-body problems,
$(\{\mathbf{\tilde{r}_k}\},\{\mathbf{R_k}\})$.  If furthermore the oscillator
frequencies and masses are related by
\begin{eqnarray} \label{cond2}
 \frac{\omega_{ik}^2}{m_{i}+m_{k}} + \frac{\omega_{i'k'}^2}{m_{i'}+m_{k'}}
 =  \frac{\omega_{i'k}^2}{m_{i'}+m_{k}} + \frac{\omega_{ik'}^2 }{m_{i}+m_{k'}}  \;,
\end{eqnarray}
the $2N$ problem separates to one $N$-body oscillator problem,
$\mathbf{R_k}$, and $N$ independent two-body problems,
$\mathbf{\tilde{r}_k}$.  This is a huge simplification in itself but
in addition also because the $N$ coupled oscillators can be solved
analytically leaving only $N$ decoupled two-body problems.  For both
conditions in eqs.~(\ref{cond1}) and (\ref{cond2}) to hold we must
have $m_{i}=m_{i'}$ for all $i$, that is all these pairs must have the
same masses.  The conditions in eqs.~(\ref{cond1}) and (\ref{cond2})
then becomes
\begin{eqnarray} \label{cond3}
  \omega_{ik}^2 + \omega_{i'k'}^2 =  \omega_{ik'}^2 + \omega_{i'k}^2     \;.  
\end{eqnarray}
One solution is then that the two frequencies between particle $i$ and
particles $k$ and $k'$ could be equal ($\omega_{ik}^2 =
\omega_{ik'}^2$), provided precisely the same holds independently for
interactions between $i'$ and particles $k$ and $k'$ ($\omega_{i'k'}^2
= \omega_{i'k}^2 $).  Other combinations are possible.  However, the
most obvious solution is that all frequencies within two layers are
equal, that is
\begin{equation} \label{freqcond}
\omega_{ik}^2 = \omega_{i'k}^2 = \omega_{ik'}^2 = \omega_{i'k'}^2 \;.
\end{equation}

We emphasize that these conditions can still be obeyed if $m_{i} \neq
m_{k}$ when $i \neq k$, provided the frequencies are related through
eq.~(\ref{cond3}).  The problem still reduces tremendously as all
coupling terms vanish.  We also want to emphasize that masses and
frequencies between different layers can be different while decoupling
still is achieved, but we shall not here pursue these possibilities.

Decoupling can still be achieved when any of the layers is only occupied by
one particle.  If $i'$ does not exist, we have to omit all terms where
it enters. For the decoupling conditions this means that the right
hand side of eq.~(\ref{cond1}) must vanish and eq.~(\ref{cond2})
reduces to the same condition.  For precisely this pair the
frequencies only have to be related by that mass weighting
where $m_{k}$ may differ from $m_{k'}$.  However, as soon as other
layers occupied by two particles are present the $m_{k} = m_{k'}$
reappears.  We are then left with eq.~(\ref{freqcond}) where the $i'$
quantities are removed.

If the simple conditions in eq.~(\ref{freqcond}) are met for
decoupling, the Hamiltonian separates into a sum of four terms
\begin{eqnarray} \label{ham0}
  H &=&   V^{(shift)} + H_{S} + H_{CM} + \sum_{k=1}^{N} H_k \;, \\ \label{hamcm}
  H_{CM} &=& -\frac{\hbar^2}{2 M_{CM}} \nabla_{\mathbf{R_{CM}}}^2 
  + \frac{1}{2} M_{CM} \omega_{CM}^2  R_{CM}^2 \;, \\ 
 H_k &=& -\frac{\hbar^2}{2\mu_{kk}}\nabla_{\mathbf{\tilde{r}}_k}^2
+V_k(\mathbf{\tilde{r}_k}) 
 \nonumber\\ \label{hampair}
 &+& \frac{1}{2}\mu_{kk} \tilde{r}_k^2 (\omega_{0k}^2 + \sum_{i=1}^{N}\frac{m_i}{m_i+m_k} \omega_{ki}^2)
\;,   \\ 
 H_{S} &=& \sum_{k=1}^{N} \left(-\frac{\hbar^2}{2M_k}\nabla_{R_k}^2
 +\frac{1}{2} M_k \omega_{0k}^2 R_k^2\right) \nonumber  \\ \label{string}
 &-&  H_{CM} +  \sum_{i=1}^{N-1}  \sum_{k=i+1}^{N}  
  2 \mu_{ik} \omega_{ik}^2 (\mathbf{R_i}-\mathbf{R_k})^2 \; . 
\end{eqnarray}
We also assume that
\begin{equation}
V^{(shift)} = 4 \sum_{i=1}^{N-1} \sum_{k=i+1}^{N} v^{(shift)}_{ik} \;,\label{shif}
\end{equation}
where the factor $4$ is from the $4$ identical interactions in
eq.~(\ref{shif}) between the particles in layers $i$ and $k$.  The center-of-mass variables are defined by
\begin{eqnarray} \label{cent1}
 M_{CM} &=&  \sum_{k=1}^N M_k \;, \\   \label{cent2}
  M_{CM}  {\mathbf{R_{CM}}}  &=& \sum_{k=1}^{N} M_k \mathbf{R_k}\;, \\ \label{cent3}
  M_{CM} \omega_{CM}^2 &=&  \sum_{k=1}^N M_k \omega_{0k}^2 \;.
\end{eqnarray}
Thus, the total center-of-mass, intra- and inter-layer coordinates are
separated.  The intra-layer coordinates are further separated into $N$
two-body systems, where their effective interaction is from the
external trap, the direct pair interaction, and from the
decoupling procedure.  The latter two terms amount to an oscillator
potential.  The inter-layer coordinates appear in the Hamiltonian as
$N$ harmonically interacting particles and the total center-of-mass
is only in a bound state if there is an external trap.

The total wave function depends initially on all coordinates, that is
$\Psi(\mathbf{r_1},\mathbf{r_2},...,\mathbf{r_N},
\mathbf{r_{1'}},\mathbf{r_{2'}},...,\mathbf{r_{N'}})$, which now
separates into products corresponding to the Hamiltonian in
eq.~(\ref{ham0}).   This simplification is written as
\begin{eqnarray} \label{wavefct}
&& \Psi(\mathbf{r_1},\mathbf{r_2},...,\mathbf{r_N},
\mathbf{r_{1'}},\mathbf{r_{2'}},...,\mathbf{r_{N'}}) \nonumber\\  
&& = \mathcal{A} \bigg(\Phi(\mathbf{R_{CM}})\bigg)
 \left(\prod_{k=1}^N\psi_k(\mathbf{\tilde{r}_k}) \right)
  \left( \prod_{i=1}^{N-1}\varphi_i(\mathbf{R'_i}) \right) \;,
\end{eqnarray}
where the Schr\"{o}dinger equations are
\begin{eqnarray} \label{scheq3}
&&  (H_{CM} - E_{CM} ) \Phi(\mathbf{R_{CM}})  = 0, \\ \label{scheq1}
&& (H_{k} - E_k)  \psi_k(\mathbf{\tilde{r}_k}) = 0,  \\
&& (H_{S} -  E_S) \prod_{i=1}^{N-1}\varphi_i(\mathbf{R'_i}) = 0 .
\label{scheq2}
\end{eqnarray}
with the $\psi_k$ functions as the solutions to the intra-layer
equations (\ref{scheq1}) from a given choice of intra-layer
interactions.  The solutions to the inter-layer equation,
eq.(\ref{scheq2}), are of the harmonic oscillator type.  The prime on
the coordinates, $R'_i$, reflects that in solving this equation,
another coordinate transformation has to be made \cite{arms11}.  The
function $\Phi(R_{CM})$ describes the harmonic center-of-mass motion.
Finally, $\mathcal{A}$ is the normalization constant and if necessary
it symbolizes symmetrization due to quantum statistics of identical
particles.  The symmetrization is only relevant for the intra-layer
wave function. 

Finally, we notice that the decoupled equations are easily found when
only one particle occupies some of the layers.  The different pieces
of the Hamiltonian as well as the related wave functions simply should
skip the corresponding summations over the primed quantities and set
$m_{i'}=0$.

\subsection{Decoupled solutions}\label{decosolve}
The solutions fall into the different parts related to intra-layer
two-body problems and inter-layer harmonic oscillator $N$-body
problems.  The two-body problems involve in general an arbitrary
interaction which therefore must be solved numerically except for
specific analytic cases.  One analytic case is obviously with
oscillator potentials which already was investigated previously even
with many particles per layer \cite{arms-dip12}.  However, repulsion cannot be
realistically simulated because an inverted oscillator would either
lead to instability or, by combination with the induced attractions
from the other layers, lead to an attractive oscillator potential.  In the next section we
shall investigate some specific more realistic analytic
intra-layer interactions.

The inter-layer problem is reduced to pure oscillator properties which
has been discussed in detail before
\cite{arms-dip12}.  If there is no one-body potential,
the center-of-mass degree of freedom solved through eqs. (\ref{hamcm})
and (\ref{scheq3}) is an unbound continuum state described by a
plane wave with the total linear momentum as the continuous quantum
number.  With an external $k$-independent oscillator trap of
frequency $\omega_{0}$, the center-of-mass frequency
given by eqs. (\ref{cent1}) and (\ref{cent3}) reduces to be
$\omega_{CM} = \omega_{0}$.

The spectrum for the inter-layer degrees-of-freedom is of oscillator
structure, that is
\begin{equation} \label{spect}
E_S(\{N_k\})= \sum_{k=1}^{N-1}\hbar\omega'_k\left(N_k+\frac{D}{2}\right),
\end{equation}
where $D$ is the spatial dimension of the layer, and $\{N_k\}$ is the
set of principal quantum numbers corresponding to each normal mode of
the solution.  The prime on the frequency, $\omega^{'}_{k}$, indicates that the $k$ solutions
may be complicated functions of the parameters in the initial
Hamiltonian \cite{arms-dip12}.  These energies then correspond to the normal modes for the
relative coordinates.  The geometry is simply described in one
dimension as an oscillating string with increasing number of nodes,
and in higher dimensions as independent product states of
corresponding strings for each dimension.  Formally the wave functions
are harmonic oscillator solutions.

Intuition for the properties of the solutions can be developed by
schematic assumptions.  We first explore the case when all one- and
two-body frequencies are the same, $\omega_{ik} \equiv \omega_{r}$ and
$\omega_{0k} = \omega_{0}$.  Two different eigenfrequencies emerge
from solving eq.~(\ref{string}), that is corresponding to
the center-of-mass motion, $\omega_{0}$, and the relative motion between the
layers, $(\omega_0^2+N \omega_r^2)^{1/2} $ \cite{arms11}. The latter
is then $N-1$ times degenerate.

Another limit is studied by ordering the coordinates, $\mathbf{R_k}$,
such that neighboring successive ``particles'' interact with the same
frequencies and all other two-body frequencies are zero.  In this
nearest neighbor approximation we then assume that
$\omega_{ik}=\omega_{12}$ for $|i-k|=1$, and with $\omega_{ik}=0$
otherwise. The solution produces $N-1$ inter-layer ``string'' modes.
The modes are all different and do not have a simple formula, but in
the large $N$ limit the smallest string frequency approaches the value
of the mean field frequency, $\omega_0$, and the highest frequency
mode becomes $\sqrt{4\omega_{12}^2+\omega_0^2}$.  The intra-layer
interaction in eq. (\ref{hampair}) also simplifies by use of the same
nearest neighbor approximation, that is
\begin{eqnarray} \label{hampair2}
H_k&=&-\frac{\hbar^2}{2\mu_{kk}}\nabla_{\mathbf{\tilde{r}}_k}^2+V_k(\mathbf{\tilde{r}})
+\frac{\mu_{kk}}{2}\left(\omega_0^2+\gamma\omega_{12}^2 \right)\tilde{r}_k^2  \;,
\end{eqnarray}
where $\gamma=1$ if $k=1$ or $N$, and $\gamma=2$ otherwise.  If we for
a moment neglect the presence of $V_k$, the frequency of the solution
is $(\omega_0^2+\gamma\omega_{12}^2)^{1/2} $.  If there is no one-body
confining field, $\omega_0=0$, then the frequency of the interior
layers is bigger than the terminal frequencies by a factor of the
square root of two.  Still a bound state solution exists. We emphasize
that the interaction, $V_k(\mathbf{\tilde{r}})$, has to be added to
obtain the proper solution.

We also need to specify how to obtain the energy shift $V^{(shift)}$
in eq.~(\ref{shif}).  The origin arises outside the present model
where the inter-layer attractive interaction is assumed to be
approximated by a two-body harmonic oscillator potential.  The
zero-point energy should in principle be chosen by using a realistic
potential to compute the two-body bound state energy, $E_{ik}$.  To
reproduce this value the oscillator must be shifted by
$v^{(shift)}_{ik}=E_{ik}-D\hbar\omega_{ik}/2$.  Without knowing the
realistic potential we can not be precise, but $E_{ik}$ should vanish
together with the energy of the oscillator.  An easy and quite natural assumption 
is linear proportionality, that is
$E_{ik} = - e_{ik} D\hbar\omega_{ik}/2$, where $e_{ik}$ for a bound
state is a positive dimensionless constant depending on the interaction
between the layers.  We then get $v^{(shift)}_{ik} = - (e_{ik}+1)
D\hbar\omega_{ik}/2$, and in total
\begin{equation} \label{shift}
V^{(shift)} = - 2 D \sum_{i=1}^{N-1} \sum_{k=i+1}^{N} (e_{ik}+1) \hbar\omega_{ik} \;,
\end{equation}
For weakly bound states $e_{ik}$ is small but the present method is much
better suited for modest or strongly bound systems. 

In summary, the total energy spectrum is a sum arising from three
different sources, that is intra- and inter-layer as well as
center-of-mass contributions. The form is therefore
\begin{eqnarray}
E &= &E_S(\{N_k\})+V^{(shift)}  \nonumber\\
&+&\hbar\omega_{CM}\left(N_{CM}+\frac{D}{2}\right)+\sum_{k=1}^{N}E_k,
\end{eqnarray}
where $N_{CM}$ is the center-of-mass quantum number, and $E_k $ is the
energy of the state within the $k$'th layer.

\section{Analytic Intralayer Solutions}\label{analytics}
The full analytic solutions of decoupled systems are determined by the
unspecified arbitrary potential, $V_k$.  It is tempting and
illuminating to continue the descriptions by assuming analytically
solvable potentials, yet with realistic or at least semi-realistic
properties.  Obviously an oscillator form,
$V_k(\mathbf{\tilde{r}})=-\mu\omega^2\tilde{r}^2/2$, allow such
solutions as studied in \cite{arms-dip12} to account for repulsion.  Other possibilities are inverse square centrifugal
potentials, $\propto 1/\tilde{r}^2$ and the
extreme short-range delta-function potential, $\propto
\delta(\mathbf{\tilde{r}})$.  These potentials are schematic
prototypes representing properties of long- and short-range character,
respectively.  We shall investigate each of these cases in the
following two subsections, working with repulsive potentials only.

\subsection{Inverse distance squared potential}

The intra-layer Hamiltonian in eq.~(\ref{hampair}) depends on the
spatial dimension.  The generic form for the inverse square potential
is
\begin{equation} \label{schrone}
H_k=-\frac{\hbar^2}{2m}\left(\frac{d^2}{dx^2} - \frac{g}{x^2}\right)
+\frac{1}{2}m\omega_k^2 x^2,
\end{equation}
where $x$ is the radial relative coordinate, $m$ is the reduced mass,
$\omega_k$ is the frequency of the oscillator term, and $g$ the
``centrifugal barrier'' strength. We note that this model is also studied 
in atomic physics where it has been dubbed the 'Crandell model' \cite{manzano2010}
follwoing Ref.~\cite{crandell1984}
The application to our system in
eq. (\ref{hampair}) is then for $D=1$ and a given $k$ achieved by
\begin{equation} \label{vari}
  m = \mu_{kk} \; \;, \;\;
  \omega_k^2 = \omega_{0k}^2 + \sum_{i=1}^{N} \omega_{ki}^2 \; \;, \;\;
 x = |\mathbf{\tilde{r}_k}| \;.
\end{equation}
The solution is known \cite{calo69,calo71}, and the quantized energies
are
\begin{equation}
E_k=\hbar\omega_k\left[2n + l_\textrm{eff} + 3/2 \right],
\label{energy}
\end{equation}
where $l_\textrm{eff} = \sqrt{g+1/4}-1/2$ and
$n$ is a non-negative integer, $n=0,1, 2,\dots$.  The label $k$ on the energy
indicates that the frequency and effective angular momentum may depend
on the layer.  The corresponding intra-layer harmonic oscillator
radial wave functions are
\begin{equation} \label{wavef}
 \psi(x)  \propto   {x}^{l_\textrm{eff}+1}\exp(-x^2/(2b_{\omega}^2)) 
L_n^{l_\textrm{eff}+1/2}(x^2/b_{\omega}^2) \;,
\end{equation}
where $x\in [0,\infty]$, $b_{\omega}^2 = \hbar/(m\omega_k)$, and
$L_n^{\alpha}(z)$ is an associated Laguerre polynomial \cite{calo69}.
The corresponding size of these states are from the same oscillator
calculations found to be
\begin{equation}
\langle x^2\rangle =  b_{\omega}^2 (2n + l_\textrm{eff} + 3/2 ) \; .
\label{radius}
\end{equation}
which through $b_{\omega}$ may depend on the layer.  

The ``layer'' may have higher spatial dimension than $D=1$. However,
the generic form in eq.~(\ref{schrone}) is still the same if we
interprete $x$ as the radial coordinate, and the Hamiltonian as
corresponding to the reduced radial equation.  The replacements necessary in
eq.~(\ref{vari}) are the same.  Now $g$ must
also include the additive centrifugal barrier contribution
arising from reduction to spherical coordinates.

For dimension, $D=2$, the centrifugal barrier term of $(l_2^2 -1)/4$
has to be added to the bare strength, $g$ where $l_2$ is an integer
related to higher partial waves.  This results in $l_\textrm{eff} =
\sqrt{g+l_2^2/4}-1/2$, the energy is given by eq.~(\ref{energy}), and
the reduced radial wave function is in eq.~(\ref{wavef}).  The total
wave function is obtained after dividing by the phase space factor,
$\sqrt{x}$, and multiplying by the angular part, $\exp(\pm i l_2
\phi)$.

Continuing to $D=3$, we have to add the ordinary centrifugal term
$l(l+1)$ to $g$, where $l$ is the angular momentum quantum number
which in turn gives $l_\textrm{eff} = \sqrt{g+(l+1/2)^2}-1/2$.  The energy in
eq.~(\ref{energy}) remains the same, and the total wave function is
found by dividing the reduced wave function in eq.~(\ref{wavef}) by
$x$, and multiplying by the angular part, that is the spherical
harmonic of order $l$ and projection, $m_l=-l,\ldots,l$.

The radial wave function is symmetric around $x=0$ for $D=1$ and
applies for bosons. For fermions we have to antisymmetrize by changing
sign for negative $x$.  For higher dimensions, application to bosons
or fermions may restrict the quantum numbers of the physically allowed
radial solutions to have the proper symmetry.

\subsection{Delta-function potential} 
We now assume that the intra-layer Schr\"{o}dinger equation from eq.~(\ref{hampair})
has $V_k$ as a zero-range potential.  This means that only
intra-layer $s$-waves are affected by this interaction where higher
partial waves remain unaltered.  We first consider dimension $D=1$,
that is
\begin{equation} \label{schrdelt}
H_k=-\frac{\hbar^2}{2m}\left(\frac{d^2}{dx^2} - \frac{2}{a_1} \delta(x)\right)
+\frac{1}{2}m\omega^2x^2  \;.
\end{equation}
where $a_1$ is the one-dimensional scattering length parametrizing the
strength of the potential.  This means repulsion for $a_1>0$ and
decreasing repulsion for increasing $a_1$. The bound-state eigenvalues, $E_k$, and
eigenfunctions, $\psi_{1D}$, for this Hamiltonian are analytically known 
\cite{busc98}. They can be found by
imposing the bound-state boundary conditions for $ x \rightarrow \pm
\infty$ combined with the appropriate limit when $x \rightarrow 0$,
that is $\partial \psi_{1D}(x)/\partial x \rightarrow
\psi_{1D}(x)/a_1$. The solutions are linear combinations of
the oscillator solutions for all non-zero $x$-values.

The energies are calculated through a transcendental eigenvalue
equation
\begin{eqnarray} \label{energyb1}
E_k = \hbar\omega_k(2n_\textrm{eff}+1/2) \; ,\;
\frac{\Gamma(-n_\textrm{eff})}{{2}\Gamma(-n_\textrm{eff}+1/2)}=\frac{a_1}{b_{\omega}} \;,\;\;
\end{eqnarray}
where $\Gamma(z)$ is the gamma function.  The related wave functions
are then
\begin{equation}
\psi_{1D}(x)\propto \exp(-x^2/(2b_{\omega}^2))U(-n_\textrm{eff},0.5,x^2/(b_{\omega}^2)),
\end{equation}
where $U(a,b,z)$ is the Tricomi confluent hypergeometric function
\cite{abr70}.  It should be noted that the values of $n_\textrm{eff}$ from eq.~(\ref{energyb1}) are
not necessarily integers.

Proceeding now to two dimensions where $V_k$ in eq.~(\ref{hampair2})
is a two-dimensional delta-function.  This requires very subtle
treatment, although the resulting equations are as simple as in one
dimension \cite{busc98,liu10}.  The oscillator potential is still
given by the frequency $\omega_k$.  The usual bound state boundary condition at infinity
has to be combined with the behavior for $x \rightarrow 0$, that is
\begin{equation} \label{schrdelt2}
 \frac{\partial \psi_{2D}(x)}{\partial x} \rightarrow 
 \frac{1}{x \ln(x/a_2)} \psi_{2D}(x),
\end{equation}
where $a_2$ is the two-dimensional scattering length, and
$\psi_{2D}(x)$ is the radial wave function
\begin{equation}
\psi_{2D}(x)\propto\exp(-x^2/(2b_{\omega}^2)) U(-n_\textrm{eff},1,x^2/b_{\omega}^2) \;.
\end{equation}
The total wave function is obtained after multiplying by an angular
function, which is just a constant since we work only with $s$-waves. The energy expression becomes
\begin{eqnarray} \label{energyb2}
E_k = \hbar\omega(2n_\textrm{eff}+1) \; ,\; 
\gamma_\textrm{E} +\frac{1}{2}\psi_{\gamma}(-n_\textrm{eff})=\ln\left(\frac{b_{\omega}}{a_2}\right) \;,\;\; 
\end{eqnarray}
where $\gamma_\textrm{E}$ is the Euler constant and $\psi_{\gamma}(-n_\textrm{eff})$ is the
digamma function \cite{abr70}.

\section{Numerical illustrations}\label{numerics}
The decoupling of inter- and intra-layer degrees-of-freedom reduces the
work to analytic oscillator problems and independent two-body problems
where specific choices for the latter can also provide fully analytical solutions.
We shall illustrate our approach with analytic results obtained from use of
both $1/x^2$ and delta-function two-body intra-layer repulsive potentials
in one and two spatial dimensions.

\subsection{Overall parameter choices}
The parameters necessary to specify the system and provide complete
solutions are masses, particle number, inter- and intra-layer
interactions, and external potential.  We assume that all particle masses are
the same, $m_k=m$, and all particles are subject to the same confinement, i.e., $\omega_{0k}=\omega_0$.  We choose energy and length
units as related to the one-body confinement potential, that is $\hbar
\omega_0$ and $\sqrt{\hbar/m\omega_0}$.  
Therefore the $m$ and $\omega_0$ dependencies are determined by the
scalings contained in these units.  The center-of-mass part
in eqs.~(\ref{hamcm}) and (\ref{cent3}) is therefore solved by the oscillator
solutions with the trap frquency, $\omega_0$.

We first choose inter-layer interactions to be only between the
nearest neighbours, which means that each of the two particles in a
layer interacts in the same way with each particle in each of the two neighbouring
layers (above and below).  Particles further apart than the adjacent layers 
are assumed not to interact.  The frequency, $\omega_{12}$, describing
this oscillator potential is chosen larger than $\omega_0$ corresponding to a
two-body attraction much stronger than the one-body potential.
We shall use the value $\omega_{12} = 3\omega_0$ in the numerical
illustrations.  The spatial dependence of the inter-layer interaction
in eq.~(\ref{string}) is then defined.

An estimate of the related total energy requires the shift in
eq.~(\ref{shift}) which in turn depends on the number of inter-layer
interacting pairs, $N-1$, and the dimension, $D$.  At the moment we
leave this two-body shift, $e_{12}$, unspecified and the total shift
is then 
\begin{equation}\label{shifteqn}
V^{(shift)} = -2D (N-1) (e_{12}+1) \hbar \omega_{12}\; . 
\end{equation}
The structure is not influenced at all by such a constant shift but it is
essential for estimates of stability against separation into smaller
clusters.

\begin{figure}
\centering
{\includegraphics[width=0.95\columnwidth]{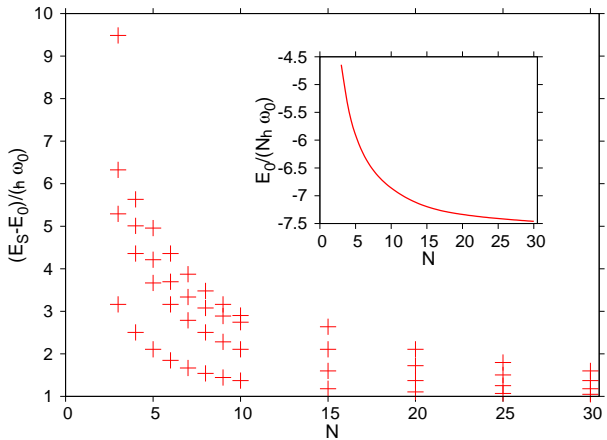}}
\caption{(Color online) Main plot: the excitation spectrum $E_S-E_0$ of
  the first four excited states of the inter-layer spectrum.
  Center-of-mass excitations are not included.  Inset: Ground state
  $E_0/N$ as function of $N$ for $D=1$.  The energy shift is included
  with $e_{12} =1$.  Only nearest-neighbor layers interact with the
  same frequency, $\omega_ {12}=3\omega_0$, all particles have equal
  mass, $m_k=m$.  The energy unit is from the external trap frequency,
  $\hbar\omega_0$.
\label{fig1}}
\end{figure}

The inter-layer and center-of-mass solutions and related structures are
now completely determined through eqs.~(\ref{string}) and
(\ref{scheq2}), and for the total energy by use of eq.~(\ref{shifteqn})
as well.  The resulting frequencies depend on $N$, as described in
Section~\ref{decosolve}.  The total energy per layer decreases for small $N$ and
saturate for large $N$ at constant values which increase with
excitation energy in accordance with the individual frequencies in
eq.~(\ref{spect}).  This saturation is closely related to the
assumption of nearest neighbor inter-layer interaction. A longer-range
interaction acting between layers further apart would lead to
more than a linear $N$-dependence of the inter-layer energy.  The
additional increase would come from the
normal mode frequencies in eq.~(\ref{spect}) which would depend 
on a higher than linear power of $N$.

The excitation energies of the lowest excited states are shown 
in fig.~\ref{fig1} along with the evolution of the ground state energies including the
zero-point shift, eq. (\ref{shifteqn}) with $e_{12}=1$, with particle number in the inset.  
As the system gets
larger, excitations become lower in energy and eventually approach the
external field frequency.  This lowering of the excitation energies
with chain length is similar to acoustic phonons in
solids \cite{arms-dip12}.  The ground state energy also decreases with each
additional layer, as each additional layer creates an additional
interior layer which has eight attractive interactions with its
neighbors.  The quantity $E/N$ approaches a constant at large $N$ since
the main contribution to the energy is linear in $N$.  These plots are
the same in 1, 2, and $3D$, if the interaction is isotropic (except for
degeneracies). The increase in zero-point energy in higher dimensions
is cancelled by the $D$-dependence of the shift in eq.~(\ref{shifteqn}).

\subsection{Inverse distance square potential}
The remaining part of the problem is related to the two-body
intra-layer Hamiltonian in eqs.~(\ref{hampair}) or (\ref{schrone}).
The fundamental frequency, $\omega$, is given by eq.~(\ref{vari}).
The intra-layer energy spectrum is given by eq.~(\ref{energy}) for
each of the $N$ layers.

The parameters of this part of the decoupled motion are layer number,
$N$, dimension, $D$, and strength, $g$, of the $1/x^2$ potential.  The
latter two only enter through $l_\textrm{eff}$, which equals $l_\textrm{eff} =
\sqrt{g+1/4}-1/2$ and $l_\textrm{eff} = \sqrt{g}-1/2$ for $s$-waves in $D=1$
and $2$, respectively.  Thus the dimension dependence is very weak and
for small $g$ only changing from $l_\textrm{eff}=0$ to $-1/2$.  The number of
layers only appears in the frequency obtained from eq.~(\ref{vari}),
but when only nearest neighbors interact, this frequency becomes
independent of the total number of layers.

The energy spectrum is then obtained by adding four spectra, that is
the inter-layer contribution shown in fig.~\ref{fig1}, the $N-2$ times
degenerate intra-layer spectrum from the inner layers, the $2$
identical end-layer spectra, and the total center-of-mass oscillator
energies.  The intra-layer contributions are $N$-independent except
for the degeneracies.  The $g$-dependence is entirely from the
intra-layer summation, where for large $g$ it approximately amounts to a shift,
$\hbar \omega \sqrt{g}$, of all the intra-layer energies.

If we also for $D=2$ include non-zero values of $l_2$, a large number
of other energies appears through the value of $l_\textrm{eff}$.  This would
for special values of $g$ corresponding to integers give rise to
precise degeneracies rather than just larger density of states.  The
$N$-dependence receives contributions from all parts of the
Hamiltonian.  The sum over all intra-layer energies is (at least for
large $N$) essentially proportional to $N$.  For large $N$ the
inter-layer energy dependence in fig.~\ref{fig1} is proportional to
$N$, and therefore comparable to the other contributing terms.  The intra-layer excitations are higher in energy than the inter-layer string excitations.  At the largest $N$ values, the intra-layer excitations approach the highest energy inter-layer excitation.

The spatial extension of the system is to a large extent determined by
the intra-layer two-body structure characterized by the mean square
radius.  These radii are given by the expression in eq.~(\ref{radius})
by insertion of the appropriate parameter values.  These harmonic
oscillator sizes are first of all determined by the length parameter,
$b_{\omega} \propto 1/\sqrt{\omega_k}$, derived from
eq.~(\ref{vari}).  The value of $\omega_k^2$ in eq.~(\ref{vari})
increases from the outer to inner layers by about a factor of $2$ for
large $\omega_{12}/\omega_0$.  These sizes are almost independent of
dimension which only enters weakly through $l_\textrm{eff}$.

\begin{figure}
\centering
{\includegraphics[width=0.95\columnwidth]{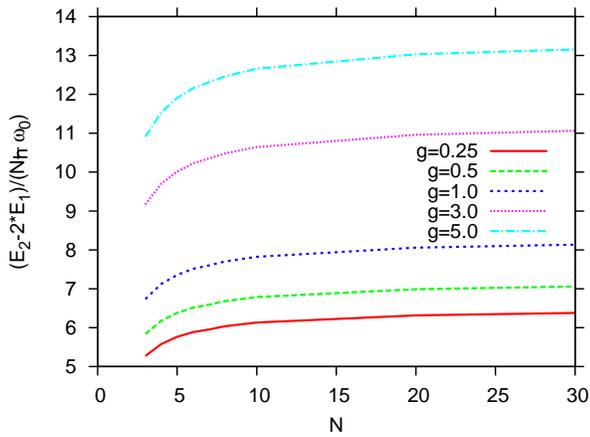}}
\caption{(Color online) The $D=1$ energies of double occupancy minus
  twice the energy of singly occupied layers as function of layer
  number $N$ for several strengths, $g$, of the $1/x^2$ potential.
  The parameters are the same as in fig.~\ref{fig1}.  We have not
  included the shift per layer, $-(e_{ik}+1) \hbar
  \omega_{12}(1-1/N)$, arising from the two-body binding energy.
\label{fig5}}
\end{figure}

The total energies are only useful in comparison with other energies.
We may view the structure as two vertical strings interacting with
each other through the direct intra-layer two-body repulsion and
indirectly through the effective intra-layer potential arising from
the interaction between particles in different strings. Then we can
investigate string-string properties where the relative binding energy
must be a crucial quantity.  We must first decide a value for
$V^{(shift)}$ to enter in both double and single string structures.
The difference between zero-point shifts for these structures becomes
$\delta V^{(shift)} = - D (N-1) (e_{12}+1) \hbar \omega_{12}$.

We show in fig.~\ref{fig5} the difference per layer between the
combined double string system and two separated strings as function of
the number of layers for different repulsive strengths.  The structure
of this energy difference can be divided into different contributions,
that is (i) inter-layer difference of doubly occupied layers minus two
times singly occupied layers, (ii) intra-layer contribution which is
almost $D$ and $N$-independent but contains all the $g$ dependence, (iii)
inter-layer center-of-mass energy difference between the two types of
occupancy which is comparably small as arising from only one (vector)
degree of freedom.

The energy difference in fig.~\ref{fig5} always increases with $N$
although approaching a constant depending on $g$ for large $N$.  The
inter-layer leading order coefficients on $N$ must be more than twice
as large for two as it is for one string with singly occupied layers.
The next to leading order vanishes after division by $N$ and the energy
difference saturates at constant asymptotic values reached within about
10\% when $N \simeq 10$.

The shifts are not included in fig.~\ref{fig5} since they strongly
depend on which inter-layer interaction we are trying to simulate.
With a shift corresponding to $e_{ik}=0$, $\delta V^{(shift)}/N
\approx -3 \hbar \omega_{0}$, the double string is never bound for any value of $N$.
This is not surprising since two particles in different layers then
have zero binding and therefore cannot supply any binding to the total
system.  If the two-body binding is given by $e_{12}=2$ all $1D$ curves in
fig.~\ref{fig5} should be translated $9$ units downwards.  Then any
number of layers are bound as long as $g$ is less than about $1$.

The $2D$ results are qualitatively very similar.  They also increase
and approach constants as function of $N$ with increasing values as
function of $g$.  The chief difference is that the $2D$ energies are
about 1.5 units of energy less than the corresponding $1D$ energy.
This occurs because the effective $g$ value is always less for $2D$
(see the discussion following eq. (\ref{energy})), so the $2D$ chain is
experiencing a weaker repulsion than the $1D$ system.  This weaker repulsion means that the larger zero-point shifts in $2D$ make the string-binding energy smaller in $2D$.

\begin{figure}
\centering
{\includegraphics[width=0.95\columnwidth]{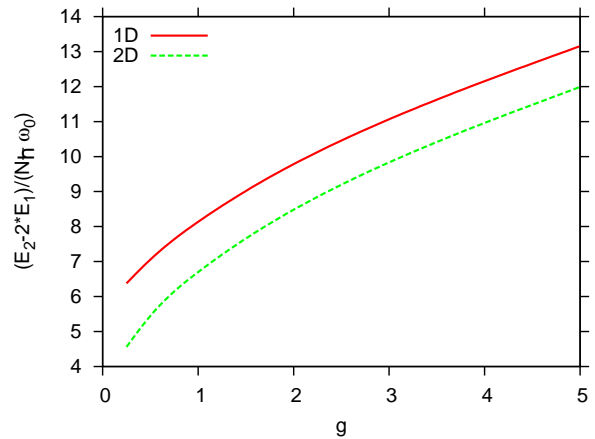}}
\caption{(Color online) The same energy differences as in
  fig.~\ref{fig5} for $N=30$, with the same parameters, but as
  function of the strength, $g$, of the $1/x^2$ potential for both
  $D=1$ and $D=2$.
\label{fig4}}
\end{figure}

The stability of a double string against two separate strings must
obviously depend on the amount of repulsion between the strings. This
arises entirely from the intra-layer $g$-dependence in the double
string energy.  We show the results for both $D=1$ and $D=2$ in
fig.~\ref{fig4} as function of $g$ for a large value of $N(=30)$ where
the asymptotic region is reached as seen in fig.~\ref{fig5}.  The
inaccurately determined shift energy is again not included but for
$\omega_{12} = 3 \omega_0$ the shift amounts to about $-3,-6,-9$ for
$D=1$ for $e_{ik}=0,1,2$, respectively, and twice these amounts for
$D=2$.  It is then easy to see at which $g$-value a large number of
layers become unstable. For fewer than $30$ layers the stability
reaches to larger values of $g$.

These results are almost totally independent of $\omega_{12}$ since
all energies for large $\omega_{12}/\omega_0$ are proportional to
$\omega_{12}$.  This includes the zero-point energies and the critical
strength, $g_{crit}$, where the separation energies are zero, therefore
remains the same.  Instability then occurs for the same
repulsion, $g$, and the same number of layers, $N$, for all large
inter-layer interaction frequencies, $\omega_{12}$.  On the other
hand, the value of $e_{12}$ is crucial for the actual estimates.

The $D=2$ results are again qualitatively very similar.  The separation
energies also increase with increasing values of $g$.  The string
separation energies in figs.~\ref{fig5} and \ref{fig4} are about $2$
energy units larger for $D=1$ than for $D=2$, that is for this choice
of $\omega_{12} = 3\omega_0$.  The reasoning here is the same as for
the previous plot, that is the centrifugal barrier present for $D=2$
effectively lowers the value of $g$.  The energy is dependent on the
square root of $g$, and that dependence is seen in the shape of the
curves in figure~\ref{fig4}.

\subsection{Delta-function intra-layer potential} 
The intra-layer extreme short-range repulsion has properties differing
from the long-range $1/x^2$ potential.  We calculate again spectra,
two-body in-layer radii, and string separation energies.  The total
spectrum has precisely the same inter-layer contribution as shown in
fig.~\ref{fig1}.  The center-of-mass term is trivially also the same.
The differences in the total spectrum are therefore contained in the
$N$ different intra-layer energies.

We therefore consider the energy expressions given in
eqs.~(\ref{energyb1}) and (\ref{energyb2}) corresponding to $D=1,2$,
respectively.  The dependence on the parameters clearly only arise
from the energy unit, $\hbar \omega_k$, and $n_\textrm{eff}$.  The frequency,
$\omega_k$, is given by eq.~(\ref{vari}) with the corresponding $N$ and
$\omega_{12}$ dependence.  The quantum number, $n_\textrm{eff}$, only depends
on the ratio between scattering length and the derived oscillator
length, $b_{\omega}$.  These spectra in both $D=1$ and $D=2$ are
discussed in \cite{busc98,liu10}.  The oscillator sequence of excited
states appears for all values of this ratio, while integer and half
integer values of the principal quantum number result for large and
small scattering lengths, respectively.

The difference to our application is only that the length unit is
$b_{\omega}$ instead of the trap length defined in eq (\ref{vari}).  This
changes the scattering lengths in these units to much smaller values,
and the asymptotic spectra are therefore reached for much smaller
$a_D/b_\omega$-values.

\begin{figure}
\centering
{\includegraphics[width=0.95\columnwidth]{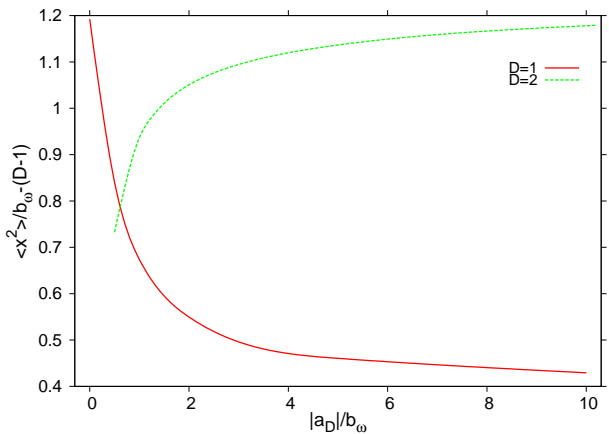}}
\caption{(Color online) The $D=1$ and $D=2$ mean square radii
  $\langle x^2\rangle/b^2_{\omega}-(D-1)$, of doubly occupied layers as function of
  strength, $a_D/b_{\omega}$, for a delta-function intra-layer
  interaction.  The other interaction parameters are the same as
  specified in fig.~\ref{fig1}.  The larger sizes for $D=2$ is
  accounted for by the shift of $(D-1)$. }
\label{fig8}
\end{figure}

The intra-layer mean square radii in units of $b_{\omega}$ are
necessarily determined by the ratio $a_D/b_{\omega}$.  The resulting
universal curves are shown in fig.~\ref{fig8} for the ground and lowest
excited states states in both $D=1$ and $D=2$.  The decreasing behavior
for $D=1$ reflects the decreasing repulsion, with the opposite behavior
for $D=2$ where the repulsion increases with increasing scattering
length. 

\begin{figure}
\centering
{\includegraphics[width=0.95\columnwidth]{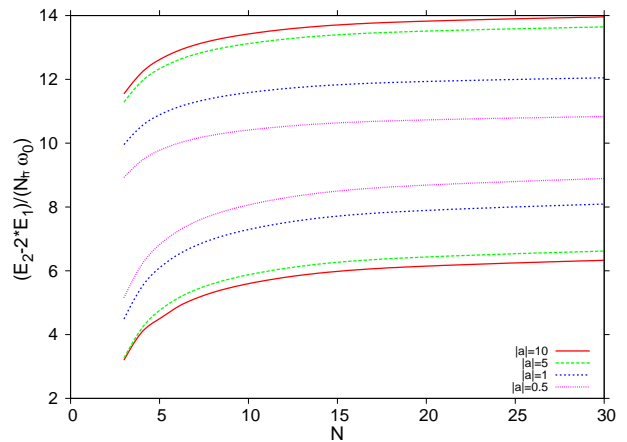}}
\caption{(Color online) The same energy differences as in
  fig.~\ref{fig5} for $D=2$ (largest values) and $D=1$ (smallest
  values) for a delta-function intra-layer interaction and several
  strengths given in terms of scattering lengths, $a_D/b_{\omega}$.
\label{fig6}}
\end{figure}

The separation energy between two coupled strings is again calculated
for the delta-function repulsion.  We show the results for $D=1$ and
$D=2$ in fig.~\ref{fig6} as function of layer number, $N$, for
different scattering lengths.  As for the $1/x^2$ potential, we find
the same behavior of an increase towards a constant large-$N$ value.
This asymptotic energy also increases with the repulsive strength.  We
have still to translate the energy scale by the shift value in order
to find critical $N$-values for given repulsion.  The results for
$D=1$ and $D=2$ are qualitatively similar, although with the order
reversed as increasing repulsion is in opposite directions of changing
scattering lengths.

\begin{figure}
\centering
{\includegraphics[width=0.95\columnwidth]{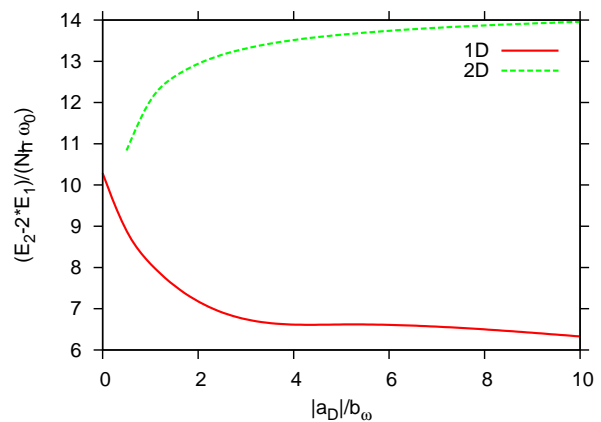}}
\caption{(Color online) The same energy differences as in
  fig.~\ref{fig4} for $N=30$ for a delta-function intra-layer
  interaction shown as functions of the strength in terms of the
  scattering lengths, $a_D/b_ {\omega}$.  Both $D=1$ (full, red) and
  $D=2$ (dashed, green) are shown.
\label{fig7}}
\end{figure}

The stability of a double string against two separate strings can be
extracted from fig.~\ref{fig7}.  The separation energy is shown as
function of decreasing repulsion for $N=30$ where the asymptotic
region is reached as seen in fig.~\ref{fig6}. This dependence arise
entirely from the intra-layer $a$-dependence in the double string
energy.  The two-body intra-layer energies reach asymptotic values for
extreme scattering lengths, which causes the many-body energies to
also flatten out.  The other contributions, including the energy
shift, are independent of the two-body repulsion and discussed in
connection with the $1/x^2$ potential.  Again, instability therefore
occurs for the same repulsion, $a_D/b_{\omega}$, and the same number of layers,
$N$, for all large inter-layer interaction frequencies, $\omega_{12}$.
The energies move in opposite directions between $D=1$ and $D=2$ given
the opposite dependence of repulsive strength and scattering length
magnitude in the different dimensions.

The scattering length dependent intra-layer excitation energies for
the delta-function potential are, like in the case of the $1/x^2$
potential, mostly higher than the string excitations.  However, in the
case of a long string in $D=1$, some of the highest string excitations
become higher in energy than the lowest intra-layer excitation.  To
give a sense of scale, the highest string excitation approaches $6.08$
$\hbar\omega_0$ when $\omega_{12}=3\omega_0$.  For a range of
scattering lengths, the lowest intra layer excitation is around $5.90$
$\hbar\omega_0$.  This excitation occurs in the outermost layers, the
interior layers' excitations are all higher than the string
excitations (coming in around 8.14 $\hbar\omega_0$).  In two
dimensions, the intra layer excitations are higher such that none of
them are lower than a string excitation, even for very long strings.

\section{Summary and conclusion}\label{conclusions}
We have formulated a prescription for a tremendous reduction of 
certain $2N$-body problems to an $N$-body oscillator problem and $N$ two-body
problems.  The conditions are rather severe but at least fulfilled for
a number of experimentally relevant one and two-dimensional systems.
The method is valid for a group of particles which can be separated
topologically in $N$ groups (layers or tubes or other separable
structures), each containing at most two arbitrarily interacting
particles.  All other one- and two-body interactions must be
modelled by harmonic oscillator potentials.

The reduction of a theoretical problem to many two-body problems is an
enormous simplification.  The Schr\"odinger equation for nearly any
interaction can be solved for two particles at least numerically, and
other observables such as momentum distributions are easily
calculated.  If these two-body spectra are calculated, then
statistical and thermodynamic quantities are also easily obtained,
since the statistical properties of oscillators are also well known.

Decoupling of the many degrees-of-freedom is only achieved for equal
masses of the two particles in each group.  Different masses are
allowed for different groups. Furthermore, two averages of the squares
of specific oscillator frequencies must be equal. These interactions
are related to the four interactions between pairs of particles in two
groups.  If one group only contains one particle it is allowed to have
an arbitrary mass, and the identical averages still hold by insertion
of zero interactions related to the removed particles.

Thus, we have one frequency constraint among four freqencies, where
the most natural solution is that all these interactions are equal.
We emphasize that this amounts to two conditions, that is equal masses
in each group, and equal interactions between two groups, but both
masses and interactions are allowed to vary between groups and pairs
of groups.  The result of these assumptions is total decoupling of
degrees-of-freedom describing the relative oscillator motion between
the center-of-masses of the groups, motion of the pairs in each layer,
and the total center-of-mass motion within the external field.  The
$(2N)$-body problem is then reduced to an analytic $N$-body oscillator
problem, $N$ independent two-body problems, and one center-of-mass
analytic oscillator problem.

We illustrate how to apply the method by calculation of a number of
basic properties in both one and two spatial dimensions.  We choose
two analytically solvable intra-group repulsive interactions where one
is the long-range inverse distance squared potential and the other is
the extreme short-range delta-function.  We first specify pertinent
analytic properties of these interactions.

Then we collect the variables describing our system, that is,
individual masses, the number of particles, the intra-group repulsive
strength, one-body external oscillator frequency, inter-group
frequency and related shift of this oscillator potential.  In the
calculations we choose as few parameters as possible while retaining a
number of fairly realistic features.  All masses and one-body
potentials are equal which leave dependence on them in the simple
scaling through a choice of length and energy units.  The inter-group
frequencies are chosen to be identical for all particles in
topologically neighboring groups and zero for all other
inter-group potentials.

We are left with one inter-group frequency, the intra-group repulsive
strengths, and the number of groups (layers or tubes).  We first
discuss the inter-group spectrum depending on $N$ and inter-group
interaction frequency. We show the saturation of these energies per
particle reached for large $N$ at values increasing linearly with
inter-group frequency.  The intra-group spectra are independent of $N$
for large $N$ and exactly given as function of repulsive strength in
the effective intra-group oscillator units arising by combination of
external trap and inter-group frequencies.  All other intra-group
properties, like root mean square radii, also only depend on the
strength in such units.

The inverse distance squared potential has very small dimensional
dependence.  The delta-function potential, characterized by scattering
length, $a_D$, leads superficially to very different behavior for
$D=1$ and $D=2$, since positive and increasing $a_1$ corresponds to
decreasing repulsion while the opposite is true for $a_2$ in two
dimensions.

The total energy is calculated as the sum of all three (inter, intra
and CM) parts.  We calculate the total energy difference between
doubly occupied groups and two times singly occupied groups.  This
energy is then the separation energy for this double string structure
into two single strings. For both intra-group repulsive interactions
and for both $D=1$ and $D=2$ we find that this energy difference
saturates at constant values for large $N$.  We calculate the
saturation values as function of the repulsive strengths.

We can then in principle predict whether these structures are stable
against string separation. For this we need to insert the energy
gained by two interacting particles in different groups.  This energy
shift would depend strongly on the initial two-body interaction,
unspecified in the present paper.  We estimate a range of values and
provide crude estimates of critical group number for stability for
different repulsion and different dimensions.

In summary, we have discussed a versatile tool to calculate
approximately a number of properties of structures separated
topologically in groups with at most two particles in each.  We
demonstrate the capacity of the method by computing various energies
as function of the variables describing the system, that is group
number, spatial dimension, and intra- and inter-group one- and
two-body interactions.  A number of practically appearring systems can
now be realistically approximated and investigated, that is for
example dipolar particles in layer and tube structures.
In the case of multiple layers, this was considered within a 
purely harmonic approach for chains with two or more particles 
per layer both at zero \cite{arms-dip12} and finite 
temperature \cite{arms-dip13}. In those studies the intralayer
interaction was harmonic and modelling repulsive intralayer
forces is therefore a delicate matter \cite{arms-dip12}. 
In comparison, the model present in the present paper allows
for different choices of the intralayer interaction other than 
harmonic. In particular, for two chains per layer one may include
the $1/r^3$ intralayer repulsion that dipoles have when their
dipoles moments are polarized perpendicular to the layer, and 
thus investigate the chain-chain dynamics. This is an interesting
topic for future study.

This work is supported by the Danish Council for Independent 
Research DFF Natural Sciences and the DFF Sapere Aude program.

\end{document}